\documentclass[11pt,a4paper]{article}
\usepackage{mcite}
\usepackage{epsfig}
\usepackage{axodraw}

\epsfverbosetrue

\def\3{\ss}

\newcommand{\tev}{{\rm Te}\kern-1.pt{\rm V}}
\newcommand{\gev}{{\rm Ge}\kern-1.pt{\rm V}}
\newcommand{\mev}{{\rm Me}\kern-1.pt{\rm V}}
\newcommand{\kev}{{\rm Ke}\kern-1.pt{\rm V}}
\newcommand{\gevsq}{\mbox{$\mathrm{{\rm Ge}\kern-1.pt{\rm V}}^2$}}
\newcommand{\gevmsq}{\mbox{$\mathrm{{\rm Ge}\kern-1.pt{\rm V}}^{-2}$}}

\begin{document}
\begin{titlepage}
%\begin{flushright}
%~\\
%\end{flushright}

\begin{center}
\begin{huge}
\bf Application of K Factors in the $H\rightarrow ZZ^\star\rightarrow 4l$ Analysis at the LHC \\
\end{huge}

\vspace{2.cm}

\Large Kyle Cranmer, Bruce Mellado, William Quayle,\\
\Large  Sau Lan Wu\\
\vspace{0.5cm}
{\Large\it Physics Department \\
University of Wisconsin - Madison \\
   Madison, Wisconsin 53706, USA }
%\maketitle

\vspace{1.5cm}

\begin{abstract}
\noindent Higher order corrections to the Higgs and the non-resonant $ZZ$
production at the LHC are evaluated within the context of the
$H\rightarrow 4l$ analysis for Higgs masses $120<M_H<180\,\gev$.
The impact of experimental cuts on the Next-to-Leading Order K factors
for Higgs and $ZZ$ production is small. The discovery potential of
the $H\rightarrow 4l$ modes is re-evaluated. With the application of
conservative higher order corrections the amount of luminosity
needed to achieve a $5\,\sigma$ signal significance with the
$H\rightarrow 4l$ modes drops by $30-35\%$, depending on the Higgs mass.

\end{abstract}
\end{center}
\setcounter{page}{0}
\thispagestyle{empty}

\end{titlepage}

\newpage

\pagenumbering{arabic}

\section{Introduction}
\label{sec:introduction}

In the Standard Model (SM) of electro-weak and strong interactions, there are 4 types of types gauge vector bosons
(gluon, photon, W and Z) and 12 types of fermions (six quarks and six
leptons)~\cite{np_22_579,*prl_19_1264,*sal_1968_bis,*pr_2_1285}.
These particles have been observed experimentally. The SM also predicts
the existence of one scalar boson, the Higgs
boson~\cite{pl_12_132,*prl_13_508,*pr_145_1156,*prl_13_321,*prl_13_585,*pr_155_1554}.
The existence of the Higgs boson remains one of the major
cornerstones  of the SM.

\begin{figure}[ht]
{\centerline{\epsfig{figure=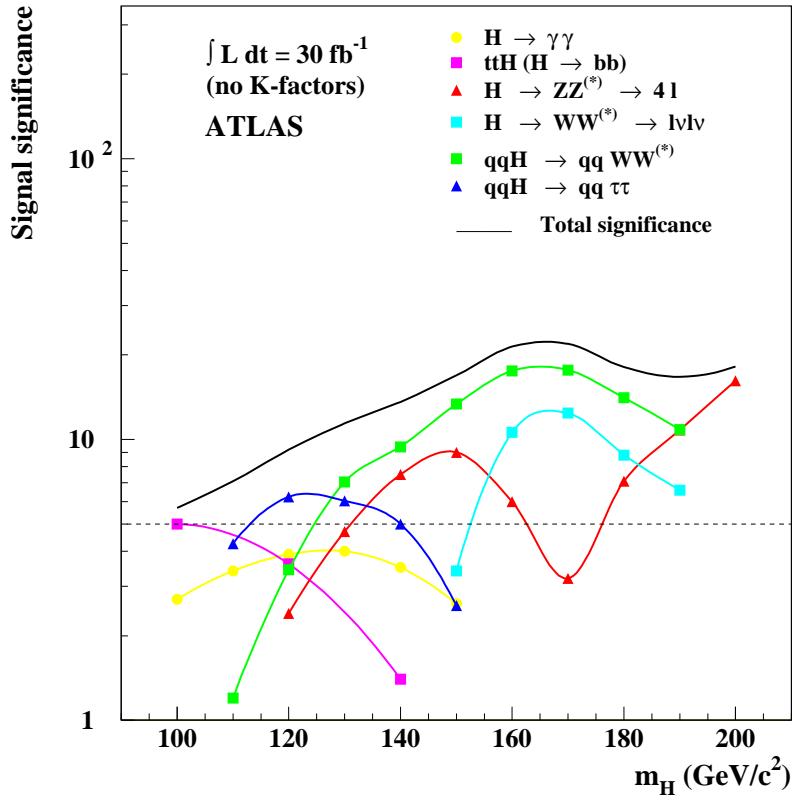,width=11.cm}}}
\caption[]{Expected Higgs signal significance obtained with the
ATLAS detector with $30\,$fb$^{-1}$ of accumulated luminosity. No K factors have been applied.}
\label{fig:vbf_signif}
\end{figure}

Nowadays, the observation of the Higgs boson is a primary focus of
the ATLAS detector~\cite{LHCC99-14}. 
The Higgs is predominantly produced at the LHC via the gluon-gluon fusion mechanism~\cite{prl_40_11_692}. The second most important production mechanism is vector boson fusion (VBF)~\cite{pl_136_196,*pl_148_367}.
The inclusion of Higgs searches using dedicated event selections
to enhance the contribution from the VBF
mechanism has dramatically enhanced the sensitivity of the LHC
experiments to low mass Higgs. Early parton level analyses showed
that VBF modes could become the most powerful discovery modes in a
large range of the Higgs mass, $M_H$,
$115<M_H<200\,\gev$~\cite{pr_160_113004,*pl_503_113,*pr_61_093005}.
More detailed analyses performed by the ATLAS collaboration which
include  initial and final state gluon radiation, hadronization,
multiple interactions and detector effects have confirmed this
statement~\cite{ATL-PHYS-2003-005} (See
Figure~\ref{fig:vbf_signif}). 

%Our group has been active  in this
%effort~\cite{ATL-PHYS-2003-002,*ATL-PHYS-2003-007,*ATL-PHYS-2003-008,*ATL-COM-PHYS-2003-002,*ATL-COM-PHYS-2003-006}.

Nevertheless, the $H\rightarrow ZZ^\star\rightarrow 4l$, or 4-lepton decay modes~\footnote{The dominant contributions arises from gluon-gluon fusion} (see the red curve with solid triangles in
Figure~\ref{fig:vbf_signif}) remain very powerful discovery modes
over a wide range of Higgs masses. The appearance of four high
transverse momentum, $P_T$, leptons providing a relatively narrow
resonance is an attractive signature experimentally. A relatively
large signal to background ratio is provided, as well.

The cross-section for the production Higgs via the gluon-gluon fusion mechanism
is subject to strong higher order corrections. The $K$ factor has been computed
to Next-to-Next-to Leading Order (NNLO) using the approximation that the top quark mass goes to infinity~\cite{prl_88_201801,np_646_220,hep-ph_0302135}. The K factor to NNLO for this production mechanism ranges between $1.7<K<2.5$, depending on the choice of factorization and renormalization scales and the Higgs mass.

Multivariate
techniques could significantly improve the signal significance by
conveniently exploiting the angular/momentum correlations between
the decay leptons. Multivariate techniques will be applied in a
future note~\footnote{In order to perform a multivariate analysis
at NLO it is mandatory to implement NLO event generators for the
signal and at least the main background processes. For the moment,
there exists an NLO event generator only for $ZZ$ production:
the MC@NLO event generator~\cite{hep-ph_02_07182}. The signal process will
be available within this MC event generator in the near
future~\cite{frixione_1}.}.

In this work we quantify the effect of including higher order corrections to signal and the
main background processes on the basis of the classical cut
analysis. The starting point of the re-evaluation of the discovery
potential of the $H\rightarrow ZZ^\star\rightarrow 4l$ modes
is the study presented in the ATLAS TDR~\cite{LHCC99-14}.

\section{A Number of Useful Definitions}
\label{sec:definitions}

A lot of effort has been invested for over a decade to estimate
QCD higher order corrections to Higgs production in collider
facilities. Next-to-Leading Order (NLO) corrections to Higgs production via gluon-gluon
fusion are known exactly~\cite{np_359_283,*pl_264_440,*prl_70_1372,np_453_17}.
NNLO calculations have been made available recently~\cite{prl_88_201801,np_646_220,hep-ph_0302135}.  These calculations were performed in the infinite top mass limit approximation~\cite{np_106_292,*sjnp_30_1368,*spu_23_429,*sjnp_44_478,np_453_17}. This
approximation is a valid one provided that the Higgs mass and the transverse momentum
of the Higgs, $P_{TH}$, are significantly smaller than the top
mass. At the LHC the Higgs is predominantly produced with
$P_{TH}<100\,\gev$. The perturbative expansion seems to be
converging for LHC energies, which gives us confidence that
N$^3$LO corrections may not play a significant role.

In recent years there has been significant progress in the
calculation of higher order corrections to the production of major
SM backgrounds in Higgs searches and the development of
NLO MC's. The major background in the the
searches for $H\rightarrow ZZ~(ZZ^\star)\rightarrow 4l$ correspond
to the non-resonant production of $ZZ (ZZ^{\star})$. The NLO
corrections to the production of weak vector bosons has been known
for over a decade and are of the order of $30\%$. One
would naively expect that NNLO corrections are not of great
relevance because of the smallness of the NLO correction. Other SM backgrounds, such as $t\overline{t}$ and $Zb\overline{b}$
production contribute little to the total background in the
searches for $H\rightarrow ZZ~(ZZ^\star)\rightarrow 4l$.

It is convenient to settle on a number of definitions in order to
avoid unnecessary confusion. In the theoretical literature the Higgs
$K(NLO)$ factor is defined as $K(NLO)=\sigma(NLO)/\sigma(LO)$,
where $\sigma(NLO)$ and $\sigma(LO)$ correspond to the total NLO
and Leading Order (LO) cross-sections, respectively, evaluated with the strong coupling constant, $\alpha_S$, and the parton density functions at the corresponding order. In general, these are
evaluated when the factorization and the renormalization scales
are set to the Higgs mass. Similarly, $K(NNLO)
=\sigma(NNLO)/\sigma(LO)$, where $\sigma(NNLO)$ is the NNLO total
cross-section.

It should be noted that in the literature the K factors sometimes
refer to the higher order corrections to the gluon-gluon fusion
mechanism, the dominant Higgs production mechanism. A
confusing notation ($pp\rightarrow HX$) is used sometimes to refer
to the higher order corrections to this
mechanism. However, for Higgs masses  $120<M_H<190\,\gev$ the
gluon-gluon fusion mechanism does not fully saturate the total
Higgs production cross-section at the LHC

Table~\ref{tab:sigcross} displays
the Leading Order (LO) Higgs cross-section production for the two major production
mechanisms at the center of mass energy of the LHC, as calculated
with the PYTHIA6.1 program~\cite{cpc_82_74,*cpc_135_238}. The contribution from the gluon-gluon fusion and the VBF mechanisms account for $\approx 70\%$ and $18-23\%$ of the total Higgs production cross-section, respectively. Additionally, the contribution from VBF to the
total cross-section is expected to increase with $P_{TH}$. The
latter is a very important discriminating variable. The
$ZZ$ system produced by the decay of the Higgs is expected to
display a significantly harder spectrum compared to that of the
non-resonant production~\cite{Mellado_10_03_03}.

\begin{table}[ht]
\begin{center}
\begin{tabular}{||c|c|c|c||c|c||}
\hline
  $M_H (\gev)$   & $\sigma_{VBF}$  & $\sigma_{gg}$ & $\sigma_{tot}$ & $\sigma_{VBF}/\sigma_{tot}$  & $\sigma_{gg}/\sigma_{tot}$  \\ \hline
120  &  4.20  &  17.21  &  23.96  &  0.18  &  0.72   \\ \hline
130  &  3.94  &  14.80  &  20.75  &  0.19  &  0.72   \\ \hline
140  &  3.61  &  13.13  &  18.28  &  0.20  &  0.72   \\ \hline
150  &  3.44  &  11.65  &  16.35  &  0.21  &  0.71  \\ \hline
160  &  3.19  &  10.46  &  14.67  &  0.22  &  0.71   \\ \hline
170  &  2.95  &  9.39   &  13.21  &  0.22  &  0.71  \\ \hline
180  &  2.80  &  8.42   &  12.04  &  0.23  &  0.71  \\ \hline
 \end{tabular}
 \caption{Values of the Higgs production Leading Order  cross-section (in pb) with the VBF mechanism, $\sigma_{VBF}$, and  the gluon-gluon fusion mechanism, $\sigma_{gg}$, calculated with  PYTHIA6.1 for different values of $M_H$ (in $\gev$). The values of $\sigma_{VBF}$ and $\sigma_{gg}$ are compared to the total Higgs cross-section, $\sigma_{tot}$. }
 \label{tab:sigcross}
\end{center}
\end{table}

In the literature higher order corrections are usually presented for the
"fully" inclusive case, after the integration over the
kinematics of the particles in the initial and final states.
However, these estimates are of little practical use except for
"guiding the eye". Experimentally, inclusive searches are
performed in a restricted fraction of the phase space. A number of
experimental criteria need to be fulfilled before the search may
be performed. Restrictions on the transverse momentum and angular
acceptance of the Higgs decay products are applied. Isolation of
leptons is crucial experimentally.

At this stage it is necessary to establish  a clear distinction between a MC integrator and an MC event generator. A MC integrator is capable of applying cuts in the phase space at parton level and also provides meaningful one dimensional distributions. This makes a MC integrator useful when calculating cross-sections in a particular region of the phase space. However, only a MC event generator is able to produce actual events, i.e., the four-momenta of incoming and outgoing particles in the hard interaction. Ideally, one would like to have NLO MC event generators to model all processes, as they can easily be interfaced with programs which model other relevant effects,  such as hadronization processes, soft and collinear QCD radiation, underlying event, multiple interactions and detector response. Today very few processes in $p-p$ collisions may be modeled with NLO event generators. LO event generators are commonly used for the estimation of the efficiencies (the ratio of the number of events reconstructed in the detector over the total number of events generated), which is a complex convolution of the kinematics of the hard interaction with a number of other effects mentioned above. Higher order corrections are usually applied ``by hand''. The effective cross-sections calculated with the help of LO event generators are scaled up with higher order corrections, or the so called K factors. Assuming that the hard interaction and the rest of the relevant effects may be factorized, these K factors may be calculated at parton level.

 In order to evaluate the NLO corrections to signal and
background after the application of experimental cuts a MC integrator
is used. This program helps evaluate the "experimental"
higher order correction which is referred here to as
$K_{CUTS}(NLO)$:
\begin{equation}
K_{CUTS}(NLO) = { \int_{CUTS}{d\sigma(NLO)\over d\Phi}d\Phi \over
\int_{CUTS}{d\sigma(LO)\over d\Phi}d\Phi  }.
 \label{eq:kcuts}
\end{equation}
The integration of the differential cross-section is performed over the phase
space, $\Phi$, as defined by the event selection performed
experimentally.

At present, no NLO event generator is available for Higgs
production at the LHC. Fortunately enough, we have at our disposal
a MC integrator that is able to apply cuts in the phase space at the
parton level.  Detector effects may not be an issue as long as the
experimental cuts are soft enough with respect to the kinematics
of the final state partons. This is the case for the inclusive
search $H\rightarrow ZZ^{\star}\rightarrow 4l$ (see
Sections~\ref{sec:tdr} and~\ref{sec:cuts}).

The MC integrator MCFM~\cite{pr_60_113006} is used for the evaluation of
$K_{CUTS}(NLO)$ for the Higgs signal~\footnote{MCFM implements the NLO ME for Higgs production using the approximation that the top quark mass goes to infinity. The cross-sections obtained  with MCFM without experimental cuts are consistent with the program HIGLU~\cite{HIGLU}.} and the $ZZ$ production.
Exact matrix elements containing the leptonic decays are
implemented in this program.

The analysis presented in the ATLAS TDR~\cite{LHCC99-14} was based
on LO matrix element (ME) interfaced with the initial and final
state radiation (IFSR), or parton shower (PS), provided within the
PYTHIA package. Strictly speaking, the resulting final state does not
correspond to  LO due to the presence of hard jets
radiated by the IFSR. The total LO cross-section is left unchanged
by the IFSR, however, the kinematics of the final state will be,
generally speaking, different  from a LO final state. The kinematics of the final state are expected to be affected little by the soft and collinear radiation.

One would like to have an event generator in which the  NLO ME is interfaced with IFSR in order to provide proper treatment of divergences
while avoiding double counting~\cite{JHEP_0206_029,np_654_301}.
Unfortunately, such type of MC event generators are still unavailable for
signal production. In order to get around that a simple procedure
is defined here to estimate the impact of hard IFSR on the kinematics
of the LO final state. The factor $\epsilon_{CUTS}(PS)$ is defined as:
\begin{equation}
\epsilon_{CUTS}(PS) = { \int_{CUTS}{d\sigma_{PS}(LO)\over d\Phi}d\Phi
\over \int_{CUTS}{d\sigma(LO)\over d\Phi}d\Phi  },
\label{eq:kcutsps}
\end{equation}
where $\sigma_{PS}(LO)$ corresponds to the cross-section obtained
with the LO ME after the application of IFSR. The integration of
the differential cross-section is performed over the phase space, as
defined by the event selection performed experimentally. The
factor $\epsilon_{CUTS}(PS)$ may be evaluated at the parton level with
the help of LO ME interfaced with IFSR provided by  PYTHIA.
This factor does not have real physical meaning by itself.
Nevertheless, it proves a useful consistency check.

As long as $\epsilon_{CUTS}(PS)$ is relatively close to unity one can
safely re-write the NLO cross-section as:
\begin{equation}
\sigma^{\star}(NLO) \approx K_{CUTS}(NLO)/\epsilon_{CUTS}(PS)\cdot
\sigma^{\star}(LO), \label{eq:kfinal1}
\end{equation}
and
\begin{equation}
\sigma^{\star}(LO) = \int_{CUTS}\left({d\sigma(LO)\over
d\Phi}\oplus PS\oplus HAD\oplus UE \oplus MI \oplus DE \right)
d\Phi, \label{eq:kfinal2}
\end{equation}
where $\sigma^{\star}(LO)$ corresponds to the LO cross-section
convoluted with the parton shower, effects of hadronization (HAD),
underlying event (UE), multiple interactions (MI) and detector
acceptance and efficiency (DE).  The TDR analysis quotes
$\sigma^{\star}(LO)$, which we are going to use as input to Equation~\ref{eq:kfinal1}.

\section{The TDR Event Selection}
\label{sec:tdr}

As pointed out in the introduction, the goal of the present note is to
re-evaluate the classical cut analysis presented in the
ATLAS TDR~\cite{LHCC99-14} by applying NLO corrections to both
signal and the major backgrounds.

The event selection applied in the ATLAS TDR~\cite{LHCC99-14}, which is based on~\cite{ATL-PHYS-1995-076,ATL-PHYS-1995-075,ATL-PHYS-1997-101}, consists of the
following cuts:
\begin{itemize}
\item At least two leptons with $P_T>20\,\gev$.
\item All leptons should have $P_T>7\,\gev$.
%\item Leptons should be isolated. The isolation radius, $\Delta
%R=\sqrt{\Delta\eta^2+\Delta\phi^2}$, is set to 0.2.
\item Leptons are required to be isolated. Stringent isolation criteria allows to suppress the $t\overline{t}$ and $Zb\overline{b}$       background below 10$\%$ of the total irreducible background~\cite{LHCC99-14}.
\item Cuts on the invariant mass of same-flavor and opposite-sign leptons. The invariant mass of one couple of leptons is required to be close to the $Z$ mass. The other couple of leptons needs to have an invariant mass larger than a certain threshold. The value of this threshold is a function of the Higgs mass~\cite{LHCC99-14}.
\end{itemize}

The signal efficiency was calculated as a function of the background rejection
with the help of a full simulation~\cite{ATL-PHYS-1997-101}. Table~\ref{tab:tdrcross} shows
the expected number of signal and background events with $30\,$fb$^{-1}$ of accumulated luminosity, as reported in the ATLAS TDR~\cite{LHCC99-14}. After the application of isolation
cuts the contribution from $t\overline{t}$ and $Zb\overline{b}$ production
(two leptons are produced via the semi-leptonic decay of the b
quarks) turns out to be negligible. Efficient suppression of the $t\overline{t}$ and $Zb\overline{b}$ background with lepton isolation selection will represent the most relevant experimental
     challenge for the signal observability with this analysis. The background is dominated by
the production of four leptons from the decay of non-resonant
$ZZ^\star$ and $ZZ$.

%\begin{table}[ht]
%\begin{center}
%\begin{tabular}{||c|c|c|c|c|c||}
%\hline
%  & 120   & 130  & 150 & 170 & 180 \\ \hline
%Signal & 0.14  & 0.38  & 0.89  & 0.25 &  0.66 \\ \hline
%$t\overline{t}$  & 0.00  & 0.00  &  0.00 & 0.00  &  0.00  \\ \hline
%$Zbb$  & 0.00  & 0.00  & 0.01  & 0.01  & 0.01  \\ \hline
%$ZZ^\star$  &  0.04  & 0.08  & 0.08 & 0.09  & 0.10  \\ \hline
%$Z\tau\tau$  & 0.00  & 0.01   &  0.01  & 0.00  &  0.00  \\ \hline
% \end{tabular}
% \caption{Effective signal and background cross-sections in the $H\rightarrow ZZ^\star\rightarrow 4l$ analysis as reported in the ATLAS TDR. Cross-sections are given in fb.}
% \label{tab:tdrcross}
%\end{center}
%\end{table}

\begin{table}[ht]
\begin{center}
\begin{tabular}{||c|c|c|c|c|c||}
\hline
  & 120   & 130  & 150 & 170 & 180 \\ \hline
Signal & 4.1  & 11.4  & 26.8  & 7.6 &  19.7 \\ \hline
$t\overline{t}$  & 0.01  & 0.02  &  0.03 & 0.02  &  0.02  \\ \hline
$Zb\overline{b}$  & 0.08  & 0.12  & 0.19  & 0.17  & 0.19  \\ \hline
$ZZ^\star$  &  1.23  & 2.27  & 2.51 & 2.83  & 2.87  \\ \hline
$ZZ\rightarrow\tau\tau ll$  & 0.13  & 0.20   &  0.25  & 0.08  &  0.02  \\ \hline
 \end{tabular}
 \caption{Expected  signal and background events with $30\,$fb$^{-1}$ of accumulated luminosity in the $H\rightarrow ZZ^\star\rightarrow 4l$ analysis, as reported in the ATLAS TDR. }
 \label{tab:tdrcross}
\end{center}
\end{table}

\section{Application of NLO Corrections} \label{sec:app}

\begin{figure}[ht]
{\centerline{\epsfig{figure=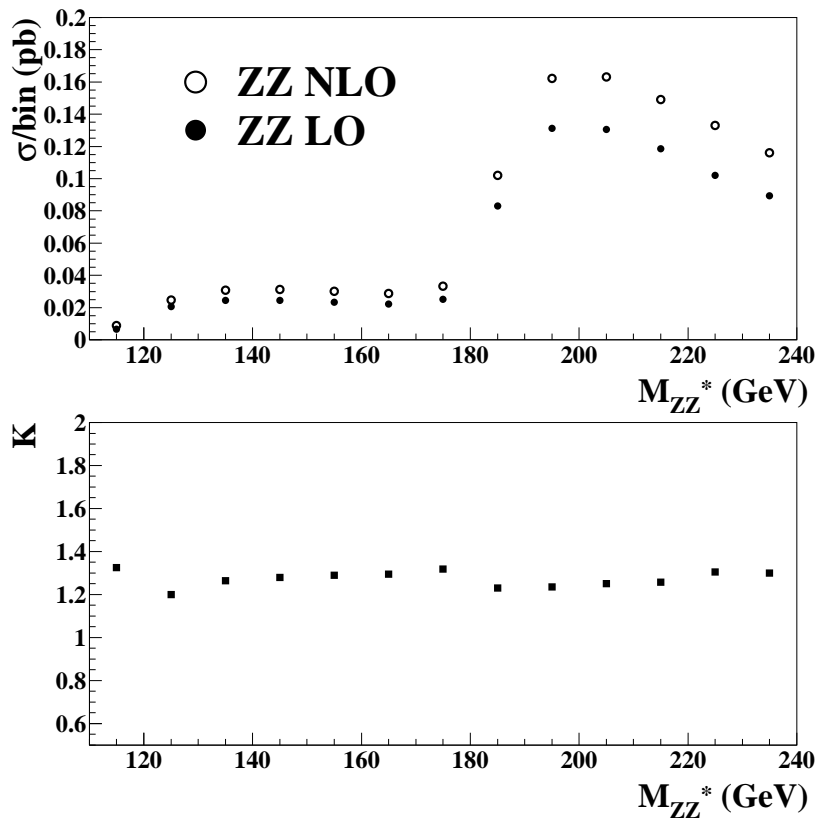,width=11.cm}}}
\caption[]{NLO corrections to non-resonant $ZZ^*$ production. The
upper plot displays the comparison between the NLO and LO
cross-sections (in pb) as a function of the invariant mass of the
vector bosons. The lower plot shows the $K(NLO)$ factor as a
function of the same variable.} \label{fig:mzz}
\end{figure}

NLO corrections are applied on the signal, $ZZ^\star$ and $ZZ$
production. The remaining background processes are scaled up by a factor
of two, in the spirit of a conservative analysis. The NLO
corrections to both processes are known, however no NLO MC event generator is
available that incorporates the corresponding NLO ME. By scaling the
$t\overline{t}$ and $Zb\overline{b}$ LO cross-sections we hope to remain on
the conservative side.

There is a great deal of  freedom in choosing the normalization of
the NLO signal cross-section. A conservative estimation of the
scale dependence of the NLO and NNLO signal cross-sections has
been performed in~\cite{prl_88_201801}. The renormalization and
factorization scales were changed in opposite directions,
yielding stronger scale dependence of the signal cross-section for the NNLO calculation. As a conservative estimate, the NLO
cross-section is chosen to match the lower bound of the NNLO
cross-section~\footnote{The lower band of the NNLO cross-sections is approximated to a straight line, $K_{gg}(NLO)=1.7+2\cdot 10^{-3}\cdot(M_H-100)$, where $M_H$ is expressed in $\gev$.}. The lower bound of the NLO band lies well outside
the NNLO band, which is supposed to contain the "true" production
cross-section. The first row in Table~\ref{tab:ksig} shows the values of the NLO corrections to the cross-section for the gluon-gluon fusion mechanism as a function of the Higgs mass.

As pointed out in Section~\ref{sec:definitions}, the gluon-gluon
mechanism does not fully saturate the total Higgs cross-section
production at the LHC. The VBF cross-section is scaled up by $10\%$ in order to take into account the NLO correction~\cite{prl_69_3274}. The rest
of the Higgs production mechanisms have not been corrected. Their
contribution to the total cross-section remains within the few
$\%$ level. The resulting values total signal $K(NLO)$ used in
this note are given in the second row of Table~\ref{tab:ksig}.

\begin{table}[ht]
\begin{center}
\begin{tabular}{||c|c|c|c|c|c||}
\hline
  & 120   & 130  & 150 & 170 & 180 \\ \hline
$K_{gg}(NLO)$ & 1.74  & 1.76  & 1.80 & 1.84 & 1.86  \\ \hline 
$K(NLO)$ & 1.47  & 1.47  & 1.50  & 1.53  &  1.54 \\ \hline
$K_{CUTS}(NLO)$ & 1.41  & 1.43  & 1.46  & 1.48  &  1.50 \\ \hline
 \end{tabular}
 \caption{Values of the signal NLO correction factors for signal production as a function of the Higgs mass. $K_{gg}(NLO)$ is the NLO correction factor to the gluon-gluon fusion production mechanism. The values of $K_{CUTS}(NLO)$ obtained with MCFM are shown in the third row (see Section~\ref{sec:cuts}).}
 \label{tab:ksig}
\end{center}
\end{table}

The analysis presented in~\cite{ATL-PHYS-1997-101} corrected the LO non-resonant $ZZ$ cross-section by applying a factor of 1.3, in order to take account of the contribution of $gg\rightarrow ZZ$. MCFM does not provide the matrix elements corresponding to the non-resonant $gg\rightarrow ZZ$ production. In this Note we apply the $K(NLO)$ factor on the non-resonant $ZZ$ cross-section presented in~\cite{ATL-PHYS-1997-101} (including the application of the factor of 1.3 mentioned above).

The $K(NLO)$  for non-resonant $ZZ^\star$ and
$ZZ$ productions are expected to be insensitive to the invariant mass of the vector
boson pair. The upper plot in
Figure~\ref{fig:mzz} displays the comparison between the NLO and
LO cross-sections as a function of the invariant mass of the
vector bosons. The lower plot shows the $K(NLO)$ factor as a
function of the same variable. Additionally, as will be seen in
Section~\ref{sec:cuts}, the effect on the NLO corrections of the cuts on the lepton transverse momentum  is small.
Therefore, it is safe to use the same $K(NLO)$ factor to correct
the LO cross-section of $ZZ\rightarrow\tau\tau ll$ production as the one used for
$ZZ^\star$ production. The scale uncertainty associated to the NLO
correction to this process due to the scale dependence is less
than $5\%$~\footnote{It should be noted
that in this check the renormalization and factorization scales
were set to be equal. This may lead to the underestimation of the
scale dependence, although this effect is expected to be small
compared with the gluon-gluon fusion Higgs production mechanism.
}.

\subsection{Effect of Event Selection on NLO Corrections}
\label{sec:cuts}

\begin{figure}[ht]
{\centerline{\epsfig{figure=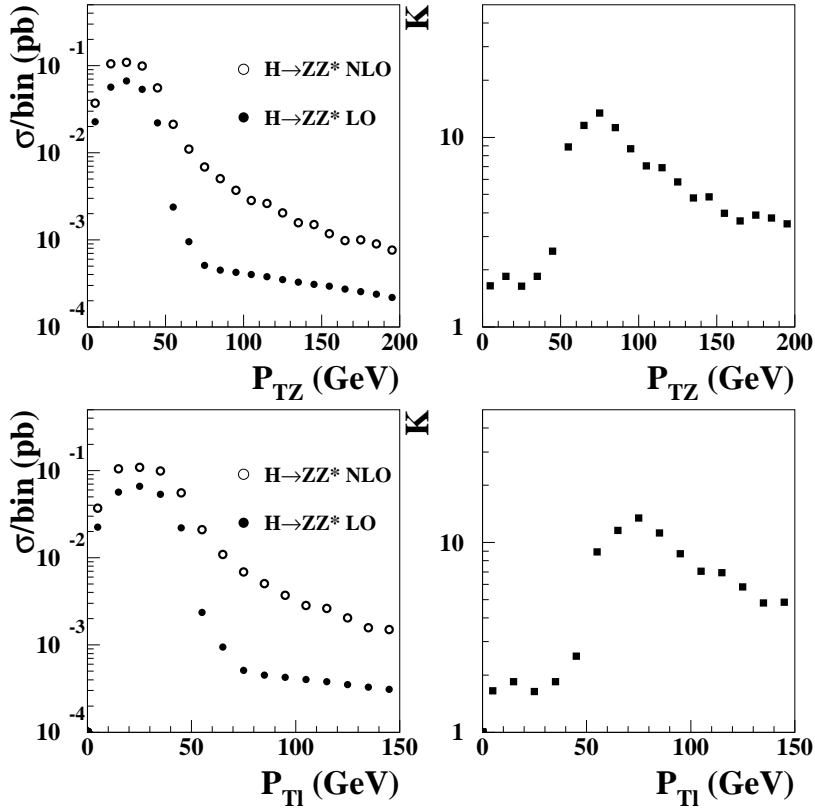,width=11.cm}}}
\caption[]{$Z(Z^\star)$ boson and lepton transverse momentum distributions with  LO and NLO $gg\rightarrow H\rightarrow ZZ^\star\rightarrow 4l$ ME's, as evaluated with MCFM. The upper left and right plots show the transverse momentum of the $Z(Z^\star)$'s (solid and open circles correspond to LO and NLO, respectively) and the corresponding $K$ factor as a function of the transverse momentum of the $Z(Z^\star)$'s. Similarly, the lower plots correspond to the transverse momentum of the leptons. Here, $M_H=150\,\gev$.}
\label{fig:ggh_ptz}
\end{figure}

\begin{figure}[ht]
{\centerline{\epsfig{figure=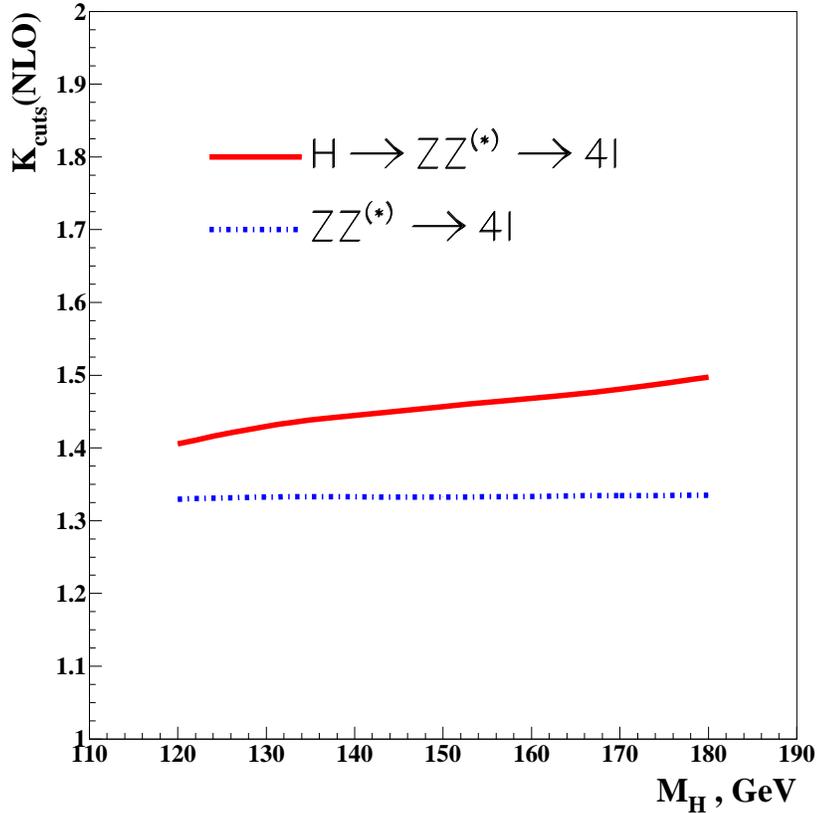,width=11.cm}}}
\caption[]{Values of $K_{CUTS}(NLO)$ for signal and the $ZZ^\star$
background are presented as a function of the Higgs mass (in the
case of the $ZZ^\star$ production, one considers the invariant
mass of the vector boson pair).} \label{fig:gghzz4l_1}
\end{figure}

\begin{figure}[ht]
{\centerline{\epsfig{figure=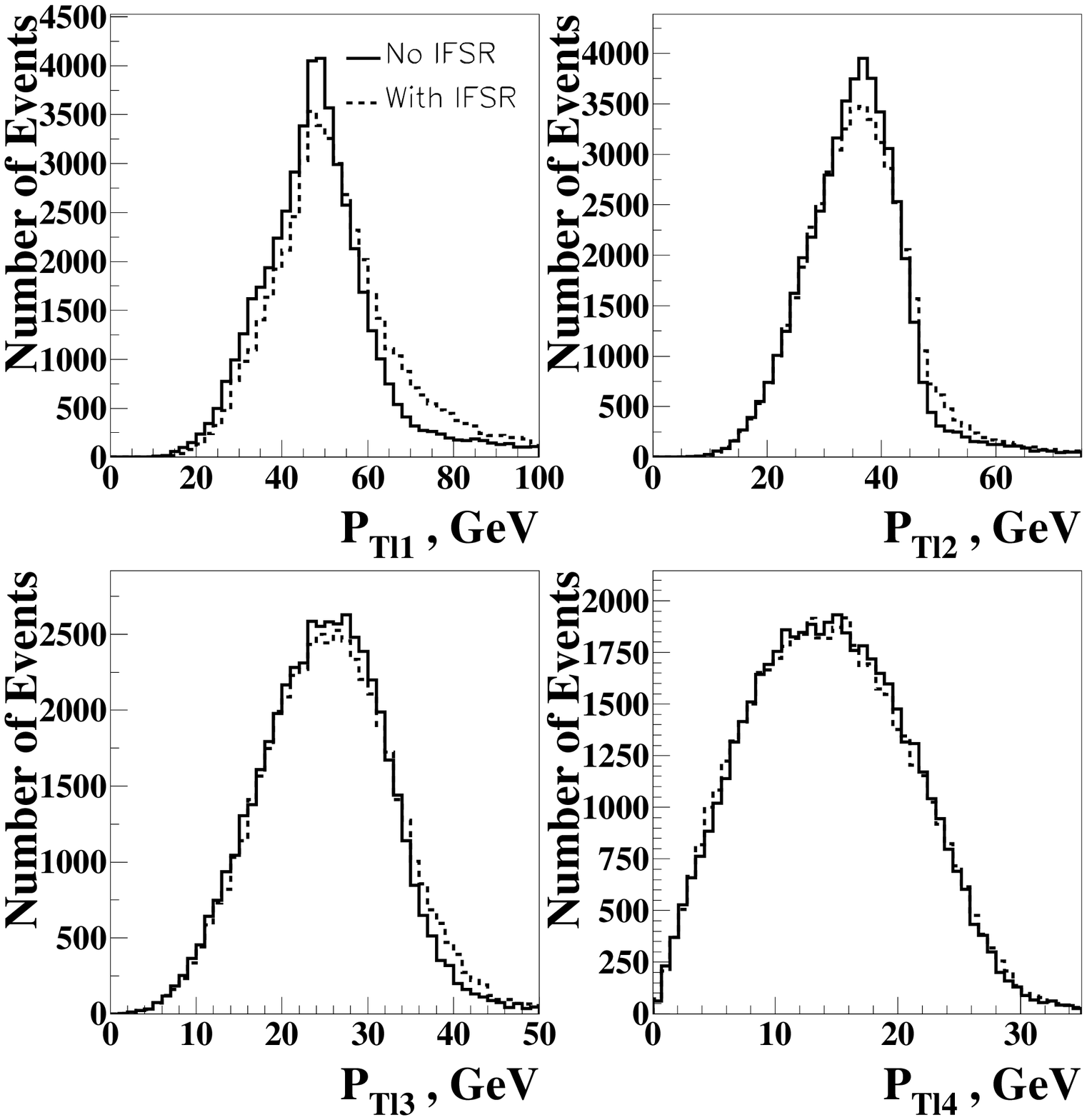,width=11.cm}}}
\caption[]{Lepton transverse momentum distribution in
$H\rightarrow ZZ^\star\rightarrow 4l$ with $M_H=150\,\gev$.
Leptons are ordered in $P_T$. The solid line corresponds to the
parton level information as produced by the LO MC without the
application of hadronization, parton showers and multiple
interaction effects. The dashed lines correspond to the same LO
MC after the application of the parton showers. Here, PYTHIA6.1 is used.}
\label{fig:hzz4l_1}
\end{figure}

As discussed in Section~\ref{sec:definitions}, the evaluation of
the effect of cuts applied experimentally on the higher order
corrections is a mandatory step. This involves the calculation of
$K_{CUTS}(NLO)$ in Equation~\ref{eq:kcuts}. This is effectively
done by the MCFM MC integrator for signal and the $ZZ^\star$
background.

Figure~\ref{fig:ggh_ptz} displays the $Z(Z^\star)$ boson and lepton transverse momentum distributions with  LO and NLO $gg\rightarrow H\rightarrow ZZ^\star\rightarrow 4l$ ME's, as evaluated with MCFM. The upper left and right plots show the transverse momentum of the $Z(Z^\star)$'s (solid and open circles correspond to LO and NLO, respectively for $M_H=150\,\gev$) and the corresponding $K$ factor as a function of the transverse momentum of the $Z(Z^\star)$'s. Similarly, the lower plots correspond to the transverse momentum of the leptons.

The kinematic effect of additional hard partons in the final state shows up at large values of the transverse momentum of the $Z(Z^\star)$'s and leptons. In the kinematic region where the cuts on the lepton transverse momentum are applied $K(NLO)$ is basically flat. Hence, the effect of experimental cuts on the NLO corrections is expected to be small for the signal process. The situation is similar for the main background processes.

The values of $K_{CUTS}(NLO)$ for signal and the $ZZ^\star$
background are presented as a function of the Higgs mass (in the
case of the $ZZ^\star$ production one considers the invariant
mass of the vector boson pair) in Figure~\ref{fig:gghzz4l_1}. The
effect of cuts is rather small in the present analysis. It reduces
the signal NLO cross-section with respect to the LO cross-section
by $\approx 5\%$. The values of $K_{CUTS}(NLO)$ in Table~\ref{tab:ksig} may be compared with $K(NLO)$ for signal. This effect in the $ZZ^\star$ background is
less than $1\%$.

\subsection{Effect of Hard IFSR}
\label{sec:ifsr}

\begin{figure}[ht]
{\centerline{\epsfig{figure=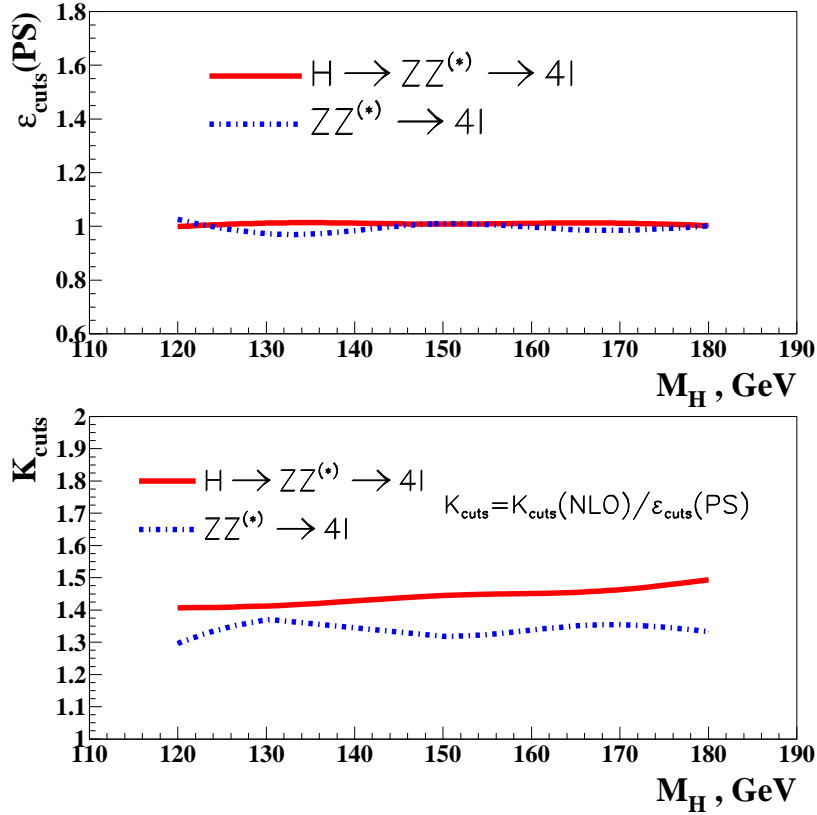,width=11.cm}}}
\caption[]{Values of $\epsilon_{CUTS}(PS)$ for signal and the $ZZ^\star$
background are presented as a function of the Higgs mass (in the
case of the $ZZ^\star$ production, one considers the invariant
mass of the vector boson pair).} \label{fig:gghzz4l_2}
\end{figure}

In the previous Section it has been shown that the presence of additional hard partons in the final state has little effect on the kinematics of relatively soft leptons. The effect of hard IFSR on the analysis reported in the ATLAS TDR may be negligible, as the parton shower approach is expected to  underestimate the rate of hard additional partons

The solid lines in Figure~\ref{fig:hzz4l_1} display the
transverse momentum distribution of the leptons, ordered in $P_T$  in $H\rightarrow ZZ^\star\rightarrow
4l$ with $M_H=150\,\gev$ (LO MC parton level information without the application of hadronization, parton showers and
multiple interaction effects). 
%The cuts applied in the TDR event
%selection are relatively soft, which is consistent with observing a small
%effect of the cuts on the NLO correction obtained in the previous Section.
The dashed lines in Figure~\ref{fig:hzz4l_1} correspond to the LO MC after the application of IFSR. The effect of hard IFSR 
is relevant to  hard leptons, however, it
affects little the $P_T$ region where the experimental cuts are
applied. The pseudorapidity distributions of the leptons are almost
unaffected by the IFSR.

The values of $\epsilon_{CUTS}(PS)$, as defined in
Equation~\ref{eq:kcutsps}, are consistent with unity within
statistical errors, both for signal and the $ZZ^\star$ background.
This is illustrated in Figure~\ref{fig:gghzz4l_2}. For practical
purposes, the final corrected cross-section in
Equation~\ref{eq:kfinal1} for signal and the $ZZ^\star$ background
are approximated by $\sigma^{\star}(NLO) \approx
K_{CUTS}(NLO)\cdot \sigma^{\star}(LO)$~\footnote{It should be noted that this conclusion does not generally apply to other inclusive analyses.}.

%~\footnote{This will not be
%case for other inclusive analyses, such as the inclusive search
%for $H\rightarrow\gamma\gamma$. In this case the cuts on the
%transverse momentum of the $\gamma$'s are significantly
%harder~\cite{Unal_12_12_02}.}.

\section{Results}

In this Section the sensitivity of the ATLAS experiment to SM
Higgs with the inclusive search with the $H\rightarrow
ZZ^\star\rightarrow 4l$ modes is re-evaluated.  The cross-sections
reported in the ATLAS TDR~\cite{LHCC99-14} have been corrected according to
Equation~\ref{eq:kfinal1} following the procedure explained in
Section~\ref{sec:app}.

Table~\ref{tab:crossresults} shows the signal and background
effective cross-section before and after the application of NLO
corrections. Shown in Table~\ref{tab:crossresults} are the signal
and background effective cross-sections, the simple event counting Poisson~\cite{ATL-PHYS-2003-008}
signal significance with 10 and $30\,$fb$^{-1}$ of accumulated
luminosity ($\sigma_P^{10}$ and $\sigma_P^{30}$, respectively) and
the amount of luminosity needed for a $5\,\sigma$ signal
significance, $L^{5\sigma}$. The last row in Table~\ref{tab:crossresults} illustrates the reduction of the $L^{5\sigma}$ due to the application of NLO corrections. The amount of luminosity needed to achieve a $5\,\sigma$ signal significance decreases by $30-35\%$, depending on the Higgs mass.

%\begin{table}[ht]
%\begin{center}
%\begin{tabular}{||c|c|c|c|c|c|c||}
%\hline
% & & 120   & 130  & 150 & 170 & 180 \\ \hline
%No K factors & $S$ &   &   &  &  &   \\ \cline{2-7}

%& $B$ &   &   & & & \\ \cline{2-7}

%& $S/B$ &   &   & & & \\ \cline{2-7}

%& $\sigma_P^{10}$ &   &   & & & \\ \cline{2-7}

%& $\sigma_P^{30}$ &   &   & & & \\ \cline{2-7}

%& $L^{5\sigma}$ &   &   & & & \\ \hline

%With K factors & $S$ &   &   &  &  &   \\ \cline{2-7}

%& $B$ &   &   & & & \\ \cline{2-7}

%& $S/B$ &   &   & & & \\ \cline{2-7}

%& $\sigma_P^{10}$ &   &   & & & \\ \cline{2-7}

%& $\sigma_P^{30}$ &   &   & & & \\ \cline{2-7}

%& $L^{5\sigma}$ &   &   & & & \\ \hline

% \end{tabular}
% \label{tab:crossresults}
%\end{center}
%\end{table}
 \begin{table}[ht]
 \begin{center}
 \begin{tabular}{||c|c|c|c|c|c|c||}
 \hline
  & & 120   & 130  & 150 & 170 & 180 \\ \hline 
 No K factors & $S$ &      0.14&      0.38&      0.89&      0.25&      0.66   \\ \cline{2-7}
 & $B$ &      0.05&      0.09&      0.10&      0.10&      0.10   \\ \cline{2-7}
 & $S/B$ &      2.83&      4.37&      8.99&      2.45&      6.35   \\ \cline{2-7}
 & $\sigma_P^{10}$ &      0.45&      2.25&      4.80&      1.25&      3.63   \\ \cline{2-7}
 & $\sigma_P^{30}$ &      2.02&      4.55&      8.88&      3.00&      6.90   \\ \cline{2-7}
 & $L^{5\sigma}$ &    124.89&     35.38&     10.68&     73.22&     16.81   \\ \hline
 With K factors & $S$ &      0.19&      0.54&      1.30&      0.38&      0.98   \\ \cline{2-7}
 & $B$ &      0.07&      0.12&      0.14&      0.14&      0.14   \\ \cline{2-7}
 & $S/B$ &      2.90&      4.56&      9.48&      2.64&      6.90   \\ \cline{2-7}
 & $\sigma_P^{10}$ &      0.94&      2.92&      6.00&      1.81&      4.70   \\ \cline{2-7}
 & $\sigma_P^{30}$ &      2.64&      5.66&     10.92&      3.86&      8.75   \\ \cline{2-7}
 & $L^{5\sigma}$ &     87.26&     24.27&      7.29&     47.39&     11.05   \\ \hline \hline
 Reduction in $L^{5\sigma}$                                  & $1-L^{5\sigma}_{WK}/L^{5\sigma}_{NK}$ &     
 0.30&      0.31&      0.32&      0.35&      0.34   \\ \hline

 \end{tabular}
 \caption{Sensitivity of the ATLAS detector to the SM Higgs with the $H\rightarrow
ZZ^\star\rightarrow 4l$ modes before and after the application of
NLO corrections as a function of the Higgs mass. Shown are the
signal and background effective cross-sections (in fb), the simple event
counting signal significance with 10 and $30\,$fb$^{-1}$ of
accumulated luminosity ($\sigma_P^{10}$ and $\sigma_P^{30}$,
respectively) and the amount of luminosity needed for a $5\,\sigma$
signal significance, $L^{5\sigma}$. The last row illustrates the reduction of the $L^{5\sigma}$ due to the application of NLO corrections ($L^{5\sigma}_{WK}, L^{5\sigma}_{NK}$ correspond to $L^{5\sigma}$ with and without the $K$ factors, respectively).}
 \label{tab:crossresults}
 \end{center}
 \end{table}

\begin{figure}[ht]
{\centerline{\epsfig{figure=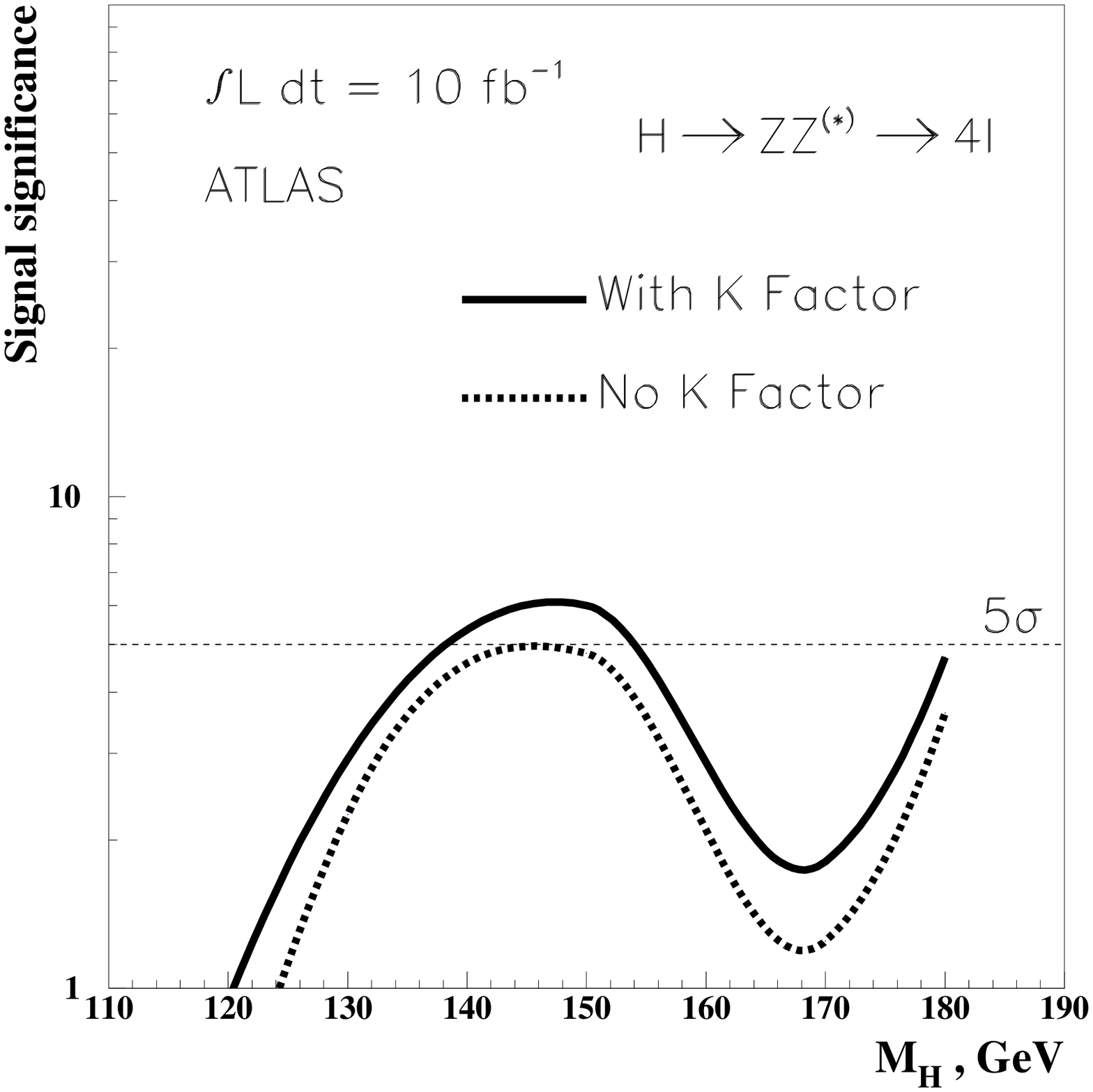,width=14.cm}}}
\caption[]{SM Higgs Signal significance expected with the ATLAS
detector with the $H\rightarrow ZZ^\star\rightarrow 4l$ modes for
 $10\,$fb$^{-1}$ of accumulated luminosity as a function of the Higgs mass. The dashed and solid curves correspond to the signal significance
before and after the application of NLO corrections,
respectively.} \label{fig:gghzz4l_4}
\end{figure}

\begin{figure}[ht]
{\centerline{\epsfig{figure=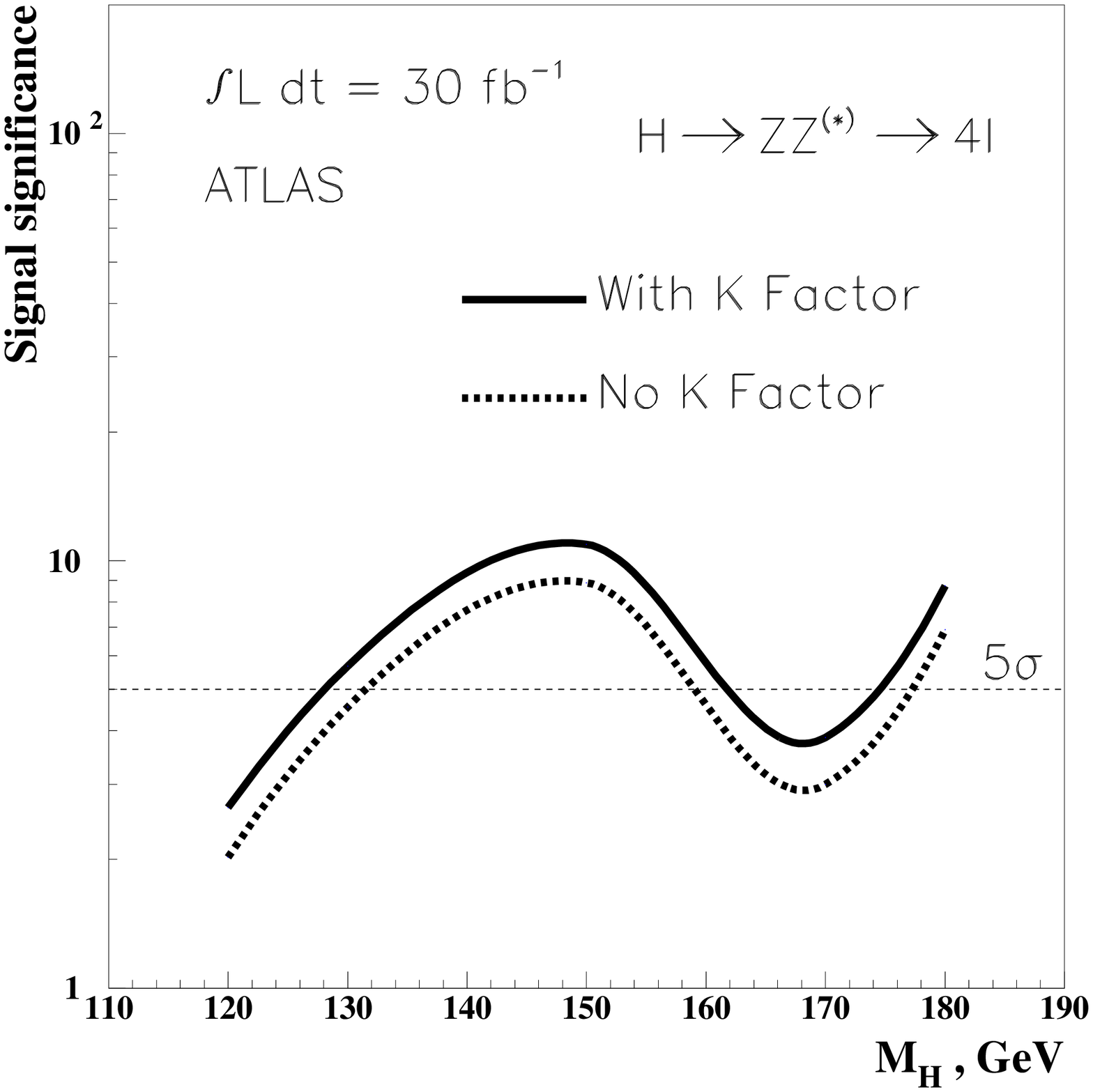,width=14.cm}}}
\caption[]{SM Higgs Signal significance expected with the ATLAS
detector with the $H\rightarrow ZZ^\star\rightarrow 4l$ modes for
 $30\,$fb$^{-1}$ of accumulated luminosity as a function of the Higgs mass. The dashed and solid curves correspond to the signal significance
before and after the application of NLO corrections,
respectively.} \label{fig:gghzz4l_5}
\end{figure}

\begin{figure}[ht]
{\centerline{\epsfig{figure=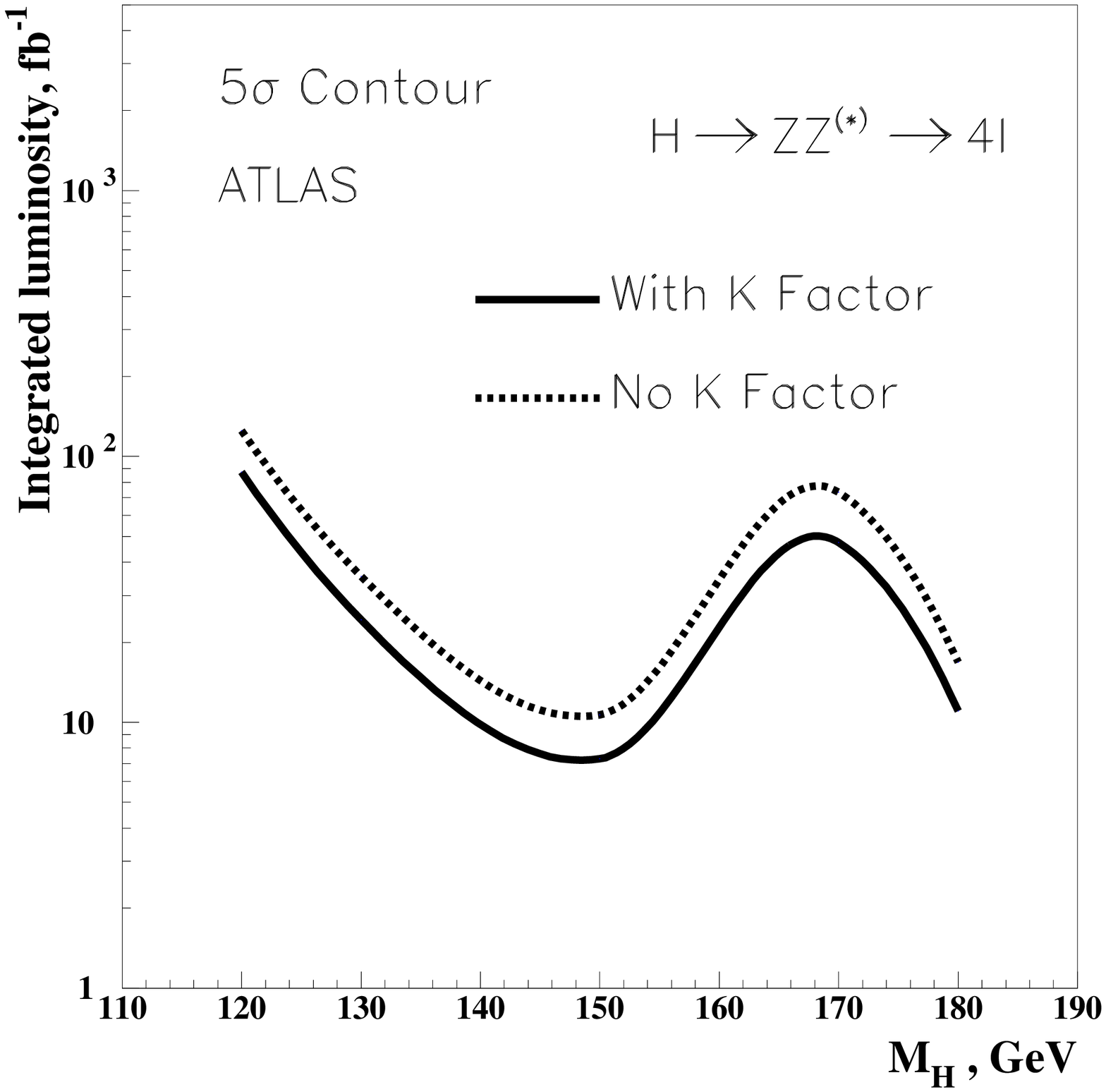,width=14.cm}}}
\caption[]{Amount of luminosity needed to achieve a $5\,\sigma$
signal significance with the ATLAS detector with the $H\rightarrow
ZZ^\star\rightarrow 4l$ modes as a function of the Higgs mass. The
dashed and solid curves correspond to the signal significance
before and after the application of NLO corrections,
respectively.} \label{fig:gghzz4l_6}
\end{figure}

Figures~\ref{fig:gghzz4l_4}-\ref{fig:gghzz4l_5} display the SM
Higgs Signal significance expected with the ATLAS detector with
the $H\rightarrow ZZ^\star\rightarrow 4l$ modes for
 $10$ and $30\,$fb$^{-1}$of accumulated luminosity, respectively.
 Figure~\ref{fig:gghzz4l_6} displays the so called luminosity
 plot. This plot shows the amount of luminosity needed to achieve a $5\,\sigma$
signal significance with the ATLAS detector with the $H\rightarrow
ZZ^\star\rightarrow 4l$ modes as a function of the Higgs mass.

A number of improvements needs to be performed in the future. The application of the NLO corrections has been performed without re-optimizing the event selection. New, more accurate, parameterizations of the background rejection and the signal efficiency will be available in the future. This will be a proper time to re-optimize the event selection.

With the appearance in the near future of NLO  event generators for the signal process it will be possible to perform more sophisticated multivariate analyses. The multivariate analyses will take advantage of the angular correlations between the leptons, not fully exploited in the past by the classical cuts analyses. 

As pointed out in Section~\ref{sec:definitions}, the transverse momentum of the four lepton system may become a powerful discriminating variable. Theoretical calculations of $P_{TH}$ based on NLO calculations with the addition of re-summation to all orders are becoming available~\cite{hep-ph_03_02104}. These calculations show that the four lepton spectrum of the signal is significantly harder than for the main background.

\section{Acknowledgments}

We would like to acknowledge J.~Campbell for most useful
discussions and his invaluable help with the MCFM program. We
would also like to thank A.~Djouadi, S.~Frixione,  M.~Grazzini, R.~Harlander, I.~Hinchliffe, E.~Richter-Was,  M.~Spira and H.~Zobernig
for most useful discussions on higher order corrections to Higgs
production at the LHC.

\bibliographystyle{zeusstylem}
\bibliography{vbf,mycites}

\end{document}